\documentclass[conference]{IEEEtran}
\IEEEoverridecommandlockouts
\usepackage{cite}
\usepackage{amsmath,amssymb,amsfonts}
\usepackage{algorithmic}
\usepackage{graphicx}
\usepackage{textcomp}
\usepackage{xcolor}
\def\BibTeX{{\rm B\kern-.05em{\sc i\kern-.025em b}\kern-.08em
    T\kern-.1667em\lower.7ex\hbox{E}\kern-.125emX}}
\usepackage{hyperref}
\begin{document}

\title{Ontology-Based Structuring and Analysis of North Macedonian Public Procurement Contracts\\

}

\author{
\centering
\IEEEauthorblockN{
    \textsuperscript{1} Bojan Ristov, 
    \textsuperscript{2} Stefan Eftimov, 
    \textsuperscript{3} Milena Trajanoska, 
    \textsuperscript{4} Dimitar Trajanov
} 
\\
\IEEEauthorblockA{
    \textit{Faculty of Computer Science and Engineering}, 
    \textit{Ss. Cyril and Methodius University}\\ 
    Skopje, North Macedonia
} 
\\
\IEEEauthorblockA{
    \textsuperscript{1} bojan.ristov@students.finki.ukim.mk, 
    \textsuperscript{2} stefan.eftimov.1@students.finki.ukim.mk, \\
    \textsuperscript{3} milena.trajanoska@finki.ukim.mk, 
    \textsuperscript{4} dimitar.trajanov@finki.ukim.mk
    }
}

% \author{
% \IEEEauthorblockN{Bojan Ristov, Stefan Eftimov, Milena Trajanoska, Dimitar Trajanov}
% \IEEEauthorblockA{\textit{Faculty of Computer Science and Engineering} \\
% \textit{Ss. Cyril and Methodius University}\\
% Skopje, North Macedonia \\
% bojan.ristov@students.finki.ukim.mk\\ stefan.eftimov.1@students.finki.ukim.mk\\
% milena.trajanoska@finki.ukim.mk\\
% dimitar.trajanov@finki.ukim.mk }
% }
% \author{
% \IEEEauthorblockN{Bojan Ristov, Stefan Eftimov, Milena Trajanoska, Dimitar Trajanov}
% \IEEEauthorblockA{\textit{Faculty of Computer Science and Engineering} \\
% \textit{Ss. Cyril and Methodius University}\\
% Skopje, Macedonia \\
% \{bojan.ristov, stefan.eftimov.1\}@students.finki.ukim.mk\\
% \{milena.trajanoska, dimitar.trajanov\}@finki.ukim.mk }
% }
% \author{
% \IEEEauthorblockN{1\textsuperscript{st} Bojan Ristov}
% \IEEEauthorblockA{\textit{Faculty of Computer Science} \\
% \textit{and Engineering}\\
% \textit{Ss. Cyril and Methodius University}\\
% Skopje, North Macedonia \\
% bojan.ristov@students.finki.ukim.mk}
% \and
% \IEEEauthorblockN{2\textsuperscript{nd} Stefan Eftimov}
% \IEEEauthorblockA{\textit{Faculty of Computer Science} \\
% \textit{and Engineering}\\
% \textit{Ss. Cyril and Methodius University}\\
% Skopje, North Macedonia \\
% stefan.eftimov.1@students.finki.ukim.mk}
% \and
% \IEEEauthorblockN{3\textsuperscript{rd} Milena Trajanoska}
% \IEEEauthorblockA{\textit{Faculty of Computer Science} \\
% \textit{and Engineering}\\
% \textit{Ss. Cyril and Methodius University}\\
% Skopje, North Macedonia \\
% milena.trajanoska@finki.ukim.mk}
% }

\maketitle

\begin{abstract}
Public procurement plays a critical role in government operations, ensuring the efficient allocation of resources and fostering economic growth. However, traditional procurement data is often stored in rigid, tabular formats, limiting its analytical potential and hindering transparency. This research presents a methodological framework for transforming structured procurement data into a semantic knowledge graph, leveraging ontological modeling and automated data transformation techniques. By integrating RDF and SPARQL-based querying, the system enhances the accessibility and interpretability of procurement records, enabling complex semantic queries and advanced analytics. Furthermore, by incorporating machine learning-driven predictive modeling, the system extends beyond conventional data analysis, offering insights into procurement trends and risk assessment. This work contributes to the broader field of public procurement intelligence by improving data transparency, supporting evidence-based decision-making, and enabling in-depth analysis of procurement activities in North Macedonia. 
\end{abstract}


\section{Introduction}
Public procurement represents a major component of government expenditure, necessitating effective management to ensure fair allocation of public funds and economic stability. However, traditional data storage formats, often tabular or unstructured, limit transparency, accessibility, and analytical depth. Addressing these challenges requires a shift toward structured, semantically enriched data representation. 

Semantic Web technologies offer a transformative solution by structuring procurement records as RDF-based knowledge graphs, enabling interconnected data, flexible querying, and advanced analytics. Ontologies define standardized vocabularies and relationships, ensuring semantic consistency and facilitating SPARQL-based queries. 

This research introduces an ontology-driven framework for procurement data analysis, integrating automated data transformation into machine-readable formats. The system leverages knowledge graphs and Machine Learning to enhance predictive modeling of procurement trends and risk factors. By combining Semantic Web technologies with data-driven analytics, this approach strengthens transparency, supports decision-making, and fosters accountability in North Macedonia’s public procurement system. 

% \section{Related Work}

% Several studies have explored the use of Semantic Web technologies in public procurement to enhance data integration and transparency. Soylu et al. (2019) introduced an ontology-based approach aligned with the Open Contracting Data Standard (OCDS), enabling structured access to procurement data. Further, the TheyBuyForYou platform (Soylu et al., 2022) leveraged linked data and modular APIs to support procurement analytics, including anomaly detection and multilingual search. Simperl et al. (2019) proposed a knowledge graph-based platform integrating procurement data across the EU, improving accessibility for various stakeholders. These efforts demonstrate the potential of ontologies and knowledge graphs to enhance procurement data management, transparency, and analytical capabilities.

\section{Data Description}

\subsection{Legal and Institutional Framework for Data Transparency}
As part of North Macedonia's commitment to enhancing transparency in public procurement, significant advancements have been made both legislatively and technologically. A key milestone in this development occurred on January 9, 2018, when the Electronic Public Procurement System (EPPS) \cite{1} was enhanced to publish awarded contracts. This improvement increased public access to procurement data, providing greater visibility into how public funds are allocated \cite{2}.

The legal foundation for this increased transparency is laid out in the Law on Public Procurement, specifically Article 6 \cite{3}, which has evolved since its initial adoption in 1998. The version adopted on February 1, 2019 (Official Gazette no. 24/2019) \cite{4}, and effective from April 1, 2019, mandates the publication of procurement notices, tender documentation, and contract awards. These legal requirements ensure that public procurement processes are conducted transparently, allowing citizens to track the allocation of public resources. 

\subsection{Data Acquisition and Source Characterization}
In alignment with these regulatory requirements, the primary data source for this study is the national open data portal \cite{5}, specifically the dedicated section for high-value contracts exceeding 1,000,000 euros \cite{6}. The procurement data published through this platform is derived from the Electronic Public Procurement System (EPPS) \cite{1}, which has been operational since 2006 and represents the longest-running procurement system in the Western Balkans region. This strategic selection enables an in-depth analysis of transactions representing substantial fiscal resource allocation within the country’s public sector. Covering a twelve-year period (2009–2021), the dataset provides a unique longitudinal perspective on procurement trends, capturing shifts influenced by political, economic, and regulatory transformations.  

\subsection{Data Format and Structure}
The raw procurement data is available in Microsoft Excel (XLSX) format, containing structured information about high-value contracts. The dataset includes essential procurement fields such as: 
\begin{enumerate}
    \item \textbf{Contracting Authority:} The institutional entity initiating the procurement process, representing various governmental strata from ministries to public enterprises and municipalities.
    
    \item \textbf{Subject of the Contract:} A textual specification delineating the procurement objective, encompassing diverse categories from infrastructure development to service acquisition.
    
    \item \textbf{Procurement Holder:} The economic operator or consortium awarded the contract, providing critical insights into public-private transactional relationships.
    
    \item \textbf{Date of the Contract:} The temporal identifier signifying formal contractual establishment, enabling chronological analysis of procurement patterns.
    
    \item \textbf{Value of the Contract in Denars:} The financial magnitude expressed in North Macedonia's national currency, providing a quantitative measurement of resource allocation.
\end{enumerate}
While these datasets provide comprehensive information, they exist as separate files requiring integration and semantic enrichment for advanced analytical capabilities. 

\section{Data Processing}
Given that procurement data is periodically published as XLSX files, an initial step involves the systematic aggregation of these individual datasets. This merging operation is conducted by identifying shared attributes across files and unifying them into a single, comprehensive dataset. 
% This consolidation process mitigates fragmentation issues and ensures the availability of a continuous and longitudinal procurement dataset spanning the designated analysis period. 

Subsequent to aggregation, the dataset undergoes an essential normalization phase. This involves removing redundant fields and standardizing column structures to ensure data consistency. Irrelevant or superfluous attributes—such as sequential numbering fields—are eliminated to refine the dataset for analytical processing. The result is a harmonized procurement database that accurately reflects high-value transactions while maintaining structural integrity.

\section{Ontology Design}

The ontology for public procurement data provides a structured model for representing key entities, their attributes, and relationships within the procurement domain. It defines fundamental procurement concepts to ensure consistency in data organization and facilitate analysis \cite{7}.

At the core of the ontology is the \textbf{Contract} class, which represents individual procurement agreements. Each contract is associated with an \textbf{Institution}, the public authority responsible for issuing it, and a \textbf{Supplier}, the economic operator fulfilling the contractual obligations. These relationships are explicitly defined through object properties:

\begin{itemize}
    \item \texttt{hasInstitution}: Links a contract to the awarding institution.
    \item \texttt{hasSupplier}: Connects a contract to the designated supplier.
\end{itemize}

In addition to defining structural relationships, the ontology includes datatype properties that capture essential contract details:

\begin{itemize}
    \item \texttt{hasAmount}: Specifies the financial value of the contract.
    \item \texttt{hasDate}: Records the issuance date of the contract.
    \item \texttt{hasDescription}: Provides a textual summary of the contract.
\end{itemize}

\begin{figure}[b]
\centering
\includegraphics[width=1\columnwidth]{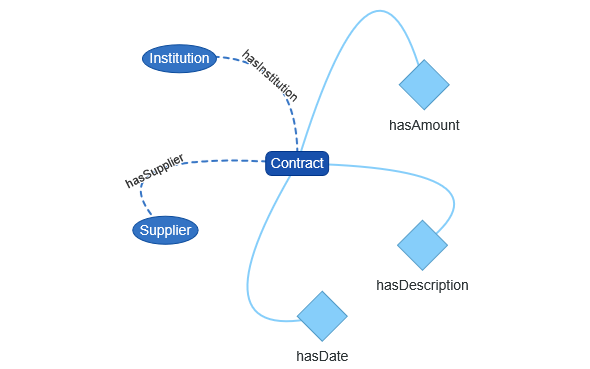}
\caption{Public Procurement Ontology Overview}
\label{fig:ontology}
\end{figure}

These properties contribute to a well-defined framework for procurement data representation (Fig. \ref{fig:ontology}).

The ontology also applies OWL (Web Ontology Language) restrictions requiring each contract to be associated with at least one institution and one supplier. This ensures the data is complete and accurately represents the procurement process \cite{8}.

\begin{figure*}[htbp]
\centering
\includegraphics[width=0.8\textwidth]{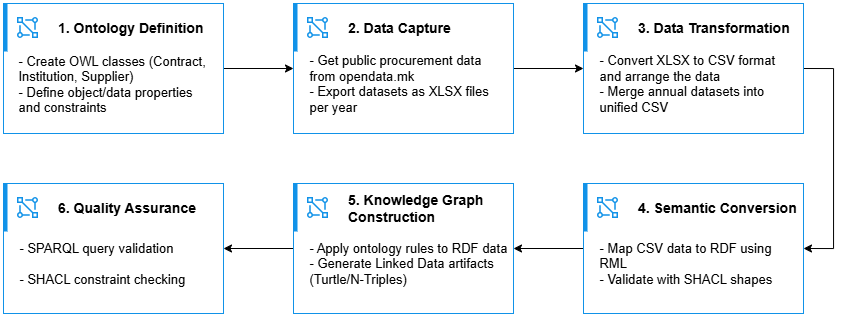}
\caption{System Workflow and Data Processing Pipeline}
\label{fig:workflow}
\end{figure*}

\section{Semantic Conversion}

To enable advanced semantic querying and Linked Open Data (LOD) compatibility, the procurement dataset is transformed from CSV to the Resource Description Framework (RDF). This conversion structures procurement records into a knowledge graph, aligning with Semantic Web standards \cite{9}.

\subsection{CSV to RDF Transformation using RML}

The transformation utilizes RDF Mapping Language (RML) to map CSV attributes to ontology properties while preserving semantic integrity \cite{9}. The process represented in Fig. \ref{fig:workflow} involves:

\begin{enumerate}
    \item \textbf{Attribute Mapping:}
    \begin{itemize}
        \item \textbf{Contracting Authority} $\rightarrow$ \texttt{Institution} class
        \item \textbf{Procurement Holder} $\rightarrow$ \texttt{Supplier} entity
        \item \textbf{Contract Value} and \textbf{Date} $\rightarrow$ \texttt{hasAmount} and \texttt{hasDate} properties
    \end{itemize}
    
    \item \textbf{Temporal Standardization:}  
    \begin{itemize}
        \item Dates are formatted in ISO 8601 to ensure uniform representation across procurement records.
        
    \end{itemize}

    \item \textbf{URI Generation:}  
    \begin{itemize}
        \item To ensure global referenceability, Unique Resource Identifiers (URIs) are systematically generated for institutions, suppliers, and contracts.
    \end{itemize}
    
    \item \textbf{RML Execution:}  
    \begin{itemize}
        \item Structured CSV data is transformed into RDF triples, forming the procurement knowledge graph.
    \end{itemize}
\end{enumerate}

The resulting RDF dataset enables procurement data to be stored in a triple store (Table \ref{tab:statistics_table}), making it accessible for SPARQL-based querying and reasoning.

\subsection{Validation with SHACL Shapes}  
After RDF transformation, the dataset undergoes validation using the Shapes Constraint Language (SHACL) to ensure compliance with ontology constraints \cite{10}.  

\begin{enumerate}  
    \item \textbf{Defining SHACL Shapes:} Each contract must have an associated Institution and Supplier. The \texttt{hasAmount} property must be a numeric value greater than zero, and \texttt{hasDate} must follow the ISO 8601 format.  
    \item \textbf{Validation Execution:} A SHACL engine checks the dataset for inconsistencies, flagging errors for correction before further analysis.  
\end{enumerate}  

\renewcommand{\arraystretch}{1.25} 
\setlength{\tabcolsep}{10pt}

\begin{table}[b]
    \centering
    \caption{Summary Statistics of the Public Procurement Dataset}

    \begin{tabular}{|l|c|}
        \hline
        \textbf{Metric} & \textbf{Count} \\
        \hline
        Contracts & 896 \\
        \hline
        Institutions & 127 \\
        \hline
        Suppliers & 563 \\
        \hline
        Total contractual links & 896 \\
        \hline
        Unique institution-supplier pairs & 662 \\
        \hline
    \end{tabular}
    \label{tab:statistics_table}
\end{table}

\section{Semantic Data Querying and Analysis in Public Procurement}

Public procurement data represents a vast and complex dataset that can be effectively managed using Semantic Web technologies. To facilitate structured querying and analysis, a knowledge graph is constructed using RDF (Resource Description Framework) and queried via SPARQL  \cite{11}.

To analyze procurement trends, a series of SPARQL queries are executed against the knowledge graph. These queries extract various key insights, such as identifying the highest contract values, calculating the total amount of all recorded contracts, analyzing the distribution of contracts over time, determining institutions with the highest number of contracts, and identifying suppliers with the greatest total contract values. Such analyses provide valuable information for monitoring public spending and detecting potential anomalies in procurement practices.

% A critical aspect of querying RDF data involves handling and transforming entity representations. The system ensures that URIs (Uniform Resource Identifiers) are decoded into human-readable labels to enhance interpretability. Additionally, numerical values such as contract amounts are formatted to maintain consistency in data representation. These preprocessing steps significantly improve the clarity and usability of extracted results.

\begin{figure}[b]
\centering
\includegraphics[width=\columnwidth]{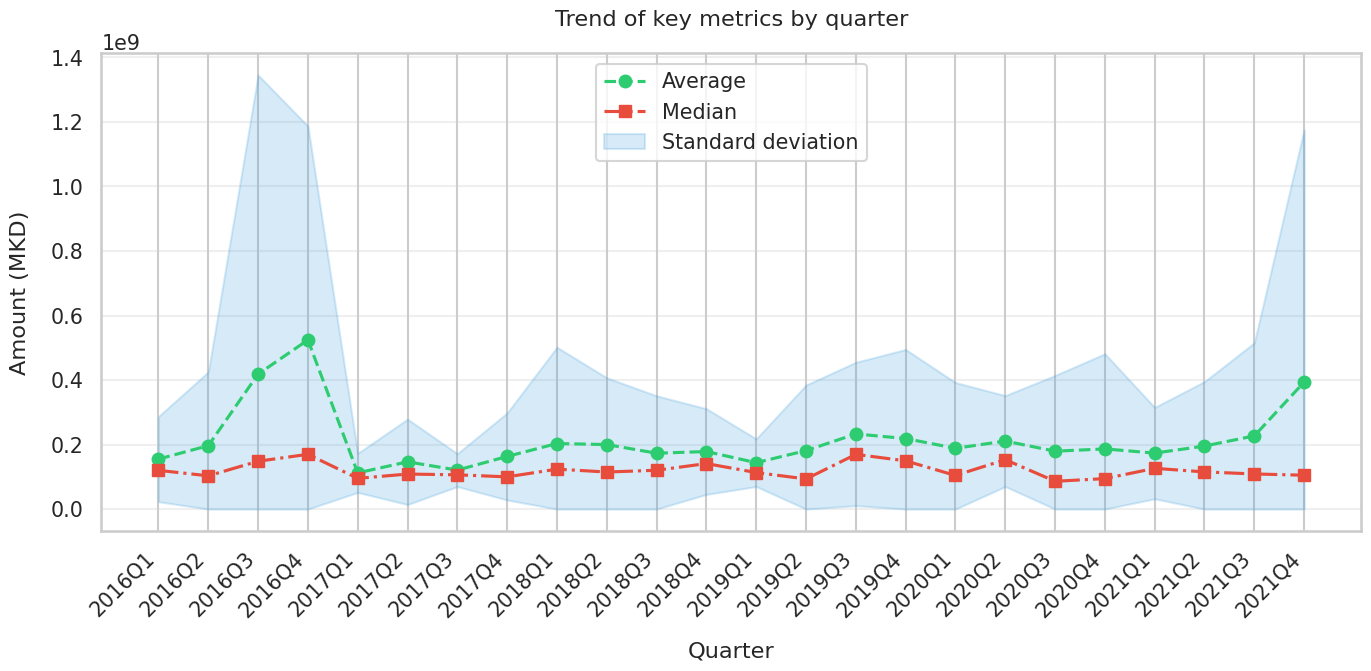}
\caption{Quarterly Trends in Public Procurement Amounts}
\label{fig:quarterly}
\end{figure}

Another key component of the analysis is the temporal distribution of contracts. By grouping procurement transactions by year and quarter, it becomes possible to identify trends in public spending over time Fig. \ref{fig:quarterly}.

Similarly, averaging contract values per month or year provides insights into fluctuations in procurement expenditures. Detecting contracts with values above the average further highlights outliers that may warrant deeper investigation. The analysis includes statistical measures such as the minimum, maximum, mean, median, and standard deviation of contract distributions for each quarter in the last five years. Part of the analysis can be seen in Table \ref{tab:sparql_queries}. 

 \renewcommand{\arraystretch}{1.12} 
 \setlength{\tabcolsep}{10pt}
\begin{table}
\centering
\caption{Key Procurement Statistics Extracted via SPARQL Queries}
\begin{tabular}{|l|c|}
\hline
\textbf{SPARQL Queries} & \textbf{Answer} \\
\hline
Total contracts & 896 contracts \\
Total amount of all contracts & 241.083.174.450 MKD \\
Year with most contracts & 2021 \\
Institution with most contracts & AD Elektrani na Makedonija \\
Highest contract value & 9.656.285.000 MKD \\
Supplier with highest total value & ALKALOID DOOEL Skopje \\
Most supplier-diverse institution & AD Elektrani na Makedonija \\
Most common pair & Ministry of health–Alkaloid \\
Average contracts per institution & 7,2205 \\
Top avg. contract value & Balkan Energy Skopje \\
Max. contracts by institution & 93 \\
Highest total (2021) & Ministry of health \\
Top contract (late December) & Balkan Energy-TE-TO Skopje \\
\hline
\end{tabular}
\label{tab:sparql_queries}
\end{table}

\section{Contract Amount Estimation and Trend Analysis}

Integrating predictive analytics into public procurement enhances decision-making by estimating contract values based on historical data and textual descriptions. The analysis is performed in two steps: procurement contract value prediction and visualization of historical spending patterns.

This strengthens transparency and analytical capabilities for data-driven procurement assessments.

\subsection{Machine Learning Model for Procurement Prediction}
We utilize the multilingual-e5-large-instruct model, a transformer-based \cite{14} sentence embedding model designed to produce high-quality vector representations of text in multiple languages. The input to the model includes the textual description of a new or unlabeled procurement contract, which is encoded into a dense vector representation.

All previously known procurement contracts from our historical dataset are pre-encoded and stored in a FAISS (Facebook AI Similarity Search) \cite{15} index to allow for efficient nearest neighbor search in high-dimensional space.

Given a new procurement contract, we compute its embedding and query the FAISS index to retrieve the top-9 most similar contracts based on cosine similarity. We then estimate the value of the new contract by computing the median of the known values of the most similar contracts.

\subsection{Model Results}
The results are displayed in the Table \ref{tab:results}. The performance of the approach is benchmarked against the median value of the amount of all contracts. 
These results indicate that the model captures key patterns in procurement data, supporting procurement planning and expenditure forecasting \cite{13}.

\renewcommand{\arraystretch}{1.1} 
\setlength{\tabcolsep}{10pt}
\begin{table}[!hb]
\centering
\caption{Results of the contract amount prediction}
\begin{tabular}{|l|l|l|l|}
\hline
\multicolumn{1}{|c|}{\textbf{Model}} & \multicolumn{1}{c|}{\textbf{RMSE}} & \multicolumn{1}{c|}{\textbf{MAE}} & \multicolumn{1}{c|}{\textbf{R2}} \\ \hline
Median (baseline)                    & 42.21                              & 13.23                             & -0.057                           \\ \hline
Embedding + FAISS                    & 39.92                              & 12.77                             & 0.056                            \\ \hline
\end{tabular}
\label{tab:results}
\end{table}

\subsection{Historical Procurement Trend Visualization}
\begin{figure}[htbp]
\centering
\includegraphics[width=\columnwidth]{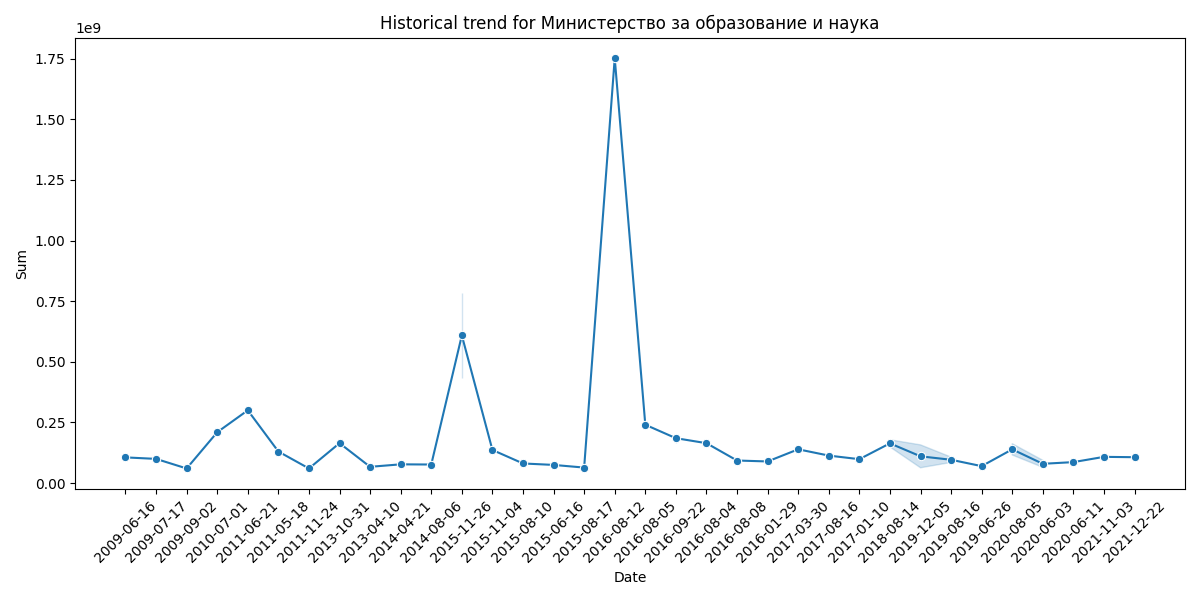}
\caption{Historical trends for Ministry of Education and Science}
\label{fig:historical}
\end{figure}

The system provides analytical tools for visualizing historical procurement trends which enable: selecting a contracting institution, retrieving and plotting past transactions and analyze spending over time.
This visualization aids in identifying spending patterns, seasonal trends, and anomalies, enabling structured procurement analysis (Fig. \ref{fig:historical}).

\section{Conclusion}

This research introduces a methodological framework for acquiring, processing, and semantically enriching North Macedonian public procurement data. By transforming tabular records into structured semantic formats, it enhances data accessibility, transparency, and analytical depth. 

% The integration of RDF and SPARQL enables efficient data exploration, supporting evidence-based decision-making and policy evaluation.

% The developed ontology provides a structured model for procurement data, facilitating integration, complex semantic queries, and advanced analytics. By incorporating ontological modeling and automated data transformation, the system enhances transparency and accountability in public procurement.

Beyond conventional analysis, the system leverages machine learning for predictive modeling and trend analysis, improving procurement forecasting while reinforcing transparency.

In conclusion, this research demonstrates how Semantic Web technologies and machine learning can transform procurement data analysis, providing stakeholders with deeper insights, anomaly detection, and improved decision-making capabilities.

\vspace{8pt}

\end{document}